 \documentclass[manuscript]{aastex}
\lefthead{}
\righthead{}
                            
\begin{document}

\title{{\it FUSE} Spectroscopy of the Transitional Magnetic Cataclysmic Variable V405 Aurigae\altaffilmark{5}}

\author{David K. Sing\altaffilmark{1}}
\author{Steve B. Howell\altaffilmark{2}}
\author{Paula Szkody\altaffilmark{3}}
\author{France A. Cordova\altaffilmark{4}}

\altaffiltext{1}{Lunar and Planetary Laboratory, Sonnett Bld., 
University of Arizona, Tucson, AZ 85721; singd@vega.lpl.arizona.edu}
\altaffiltext{2}{WIYN Observatory \& NOAO, P.O. Box 26732, 950 N. Cherry Ave., 
Tucson, AZ 85726-6732; howell@noao.edu}
\altaffiltext{3}{Department of Astronomy, University of Washington, Box 351580, Seattle, WA 98195;
szkody@astro.washington.edu}
\altaffiltext{4}{Institute of Geophysics and Planetary Physics, Department of Physics, University 
of California, Riverside, CA 92521, USA}
\altaffiltext{5}{Based on observations made with the NASA-CNES-CSA Far Ultraviolet
 Spectroscopic Explorer. FUSE is operated for NASA by the Johns Hopkins University
 under NASA contract NAS5-32985.}

\begin{abstract}
We present {\it Far Ultraviolet Spectroscopic Explorer} observations of the magnetic 
cataclysmic variable V405 Aurigae.  Together with four other DQ Her type binaries, V405 
Aur forms a small subclass of intermediate polars which are likely to evolve into low 
magnetic field strength polars, similar to AM Her.  Our co-added {\it FUSE} spectrum exhibits broad 
O VI and C III emission-lines as well as a narrow O VI emission-line component which likely forms near 
the white dwarf surface.  The O VI $\lambda1032/\lambda1038$ narrow emission line flux ratio is near 2.0 throughout the orbit, indicative of its formation in 
an optically thin gas.  Radial velocity measurements of the narrow O VI emission-lines restrict any 
orbital modulation to a very low amplitude (K$_{wd} = $2.5 $\pm$ 0.5 km s$^{-1}$) indicating, along with other 
single peaked broad disk emission lines in the UV and optical, that the binary system is at low inclination.

\end{abstract}


\newpage

\section{INTRODUCTION}
	Cataclysmic variables (CVs) are close binary star systems consisting of a white 
dwarf (WD) primary and a late type secondary which overflows its Roche lobe, often 
forming an accretion disk.  Magnetic CVs make up a particular subclass in which the 
magnetic field of the WD influences the accretion flow from its companion.  Magnetic 
CVs have two distinct subclasses, polars (or AM Her binaries) and intermediate polars 
(IPs or DQ Her binaries).  

In polars, the white dwarf rotates synchronously with its 
companion and possesses a sufficiently strong magnetic field ($\sim$ 10-100 MG) to 
prevent an accretion disk from forming. The mass transfer stream is instead funneled into 
accretion columns at one or both magnetic poles.  In intermediate polars, the WD 
rotates asynchronously with its companion, developing a partial accretion disk before the 
material is channeled onto the white dwarf.  The partially collimated material in IPs falls onto the 
magnetic poles of the WD, usually releasing its kinetic energy via thermal bremsstrahlung in the 
hard X-ray through the UV range.  Modulation of the X-ray and EUV spectrum can be attributed to 
self occultation of the WD and/or photo-electric absorption.
It is thought that as IPs evolve to shorter orbital periods, the WD's magnetic field will 
eventually synchronize its rotation with the orbit and the system will become a polar (Howell et al. 2001).

V405 Aurigae was discovered in the ROSAT all-sky survey by Haberl et al. (1994) 
and classified as an IP.  Follow up X-ray observations and optical 
spectroscopy revealed the binary to have a 4.15 hour orbital period and a 545.455 sec 
WD spin period (Table 1).  V405 Aur along with PQ Gem, RXJ0512, and RXJ0757 
(James et al. 2002, Burwitz et al. 1996, and Kemp 2002 respectively) 
form a particular subclass of IPs which are likely to soon become polars as they evolve.  
Evolutionary models show that these four binary systems have likely only recently become CVs and will probably 
become short period AM Her type stars within $\sim$1 Gyr. (Howell et al. 2001).  This group 
displays properties common to both IPs and polars.  In common with IPs this group 
displays: (1) a white dwarf spin period much shorter than the orbital period, (2) a strong, 
spin modulated X-ray pulsation, and, (3) a synodic beat period seen in optical 
observations.  The group's polar properties include: (1) spin modulated polarization, (2) 
red flux variations modulated on the rotation period, (3) a luminous cyclotron component 
indicative of stronger than normal IP magnetic field strengths, (4) a strong soft X-ray/EUV 
component similar to many high field strength polars, and (5) narrow X-ray 
dips, suggesting a highly magnetically collimated geometry of the in-falling accretion 
material.  The study of these few objects will ultimately provide many key insights into 
the accretion dynamics, magnetic field distribution on the WD surface, and evolutionary 
aspects of magnetic CVs.  

In this paper, we detail far ultraviolet spectroscopic observations of V405 
Aur and use these, as well as other existing ultraviolet and optical spectroscopic data, to 
constrain the parameters and geometry of this unique binary system.

\section{{\it FUSE} Observations}
Far Ultraviolet spectra (905-1195 \AA) of V405 Aur were obtained with the {\it Far 
Ultraviolet Spectroscopic Explorer} ({\it FUSE}) satellite (Moos et al. 2000) during Cycle 4 
observations (Table 2).  A total of 20 exposures were taken 
through the large (LWRS, 30'' $\times$ 30'') aperture in time tagged (TTG) mode 
between MJD 2452913.08758 and 2452913.59935 for a total exposure time of 22,800 s.  Data 
from the 2003 observations were processed with the {\it FUSE} data reduction software 
(CalFUSE ver. 2.4.1 and ver. 3.0.6).  The extracted spectra consist of four different 
instrument channels (separate optical paths with different mirrors and gratings) and two 
different instrumental sides (independent micro-channel plate detectors).  The extracted 
spectra, excluding the LiF 1B channel containing ``the worm'', were concatenated and co-added to produce a spectrum with the highest possible signal-to-noise for line identification purposes.  In order to observe possible O VI 
and other line variations on the white dwarf spin and orbital period, we re-extracted the time-tagged spectra as 
a function of both phases.

\section{{\it FUSE} Spectrum of V405 Aur}
      The UV spectra of V405 Aur (Figure 1) lack 
continuum emission (the error in the continuum flux is larger than the measured flux) but contain broad O VI and C III 
emission-lines which can be attributed to the accretion disk (see below).  
While a broad O VI $\lambda$1032 emission line appears in the spectra, a broad O VI $\lambda$1038 emission line does not
seem to be present and is probably too faint to detect.  The small emission lines surrounding the narrow O VI $\lambda$1038 line
are identified as C II $\lambda$1036.337, C II $\lambda$1037.018 and O I $\lambda$1039.230.
The spectra also show a narrow emission-line component for the O VI resonance 
doublet $\lambda\lambda$1032, 1038 (Figure 2); but no He II $\lambda$1085 emission is evident.  The other {\it FUSE} spectral 
lines, shown in Figures 1 and 2, were identified as airglow.  The emission-line FUV spectra 
of V405 Aur closely resembles that seen in AM Her (Hutchings et al. 2002; Mauche \& Raymond 1998), 
having both narrow and broad 
emission-line components, although AM Her does not contain an accretion disk.  The FWHM of the 
broad emission-lines seen in the {\it FUSE} spectrum of V405 Aur measure 4.5 \AA\ corresponding 
to a disk broadening of 1100$\pm$50 km s$^{-1}$ for 
the C III $\lambda$ 1175 line.  

Measurements of the O VI narrow emission-lines were performed in the 
Interactive Data Language (IDL) software package, recording empirical parameters, such as equivalent width, as 
well as those found from fitting the line to a Gaussian profile, which were used for measuring radial velocities.  An eleven 
point, $\sim$0.06 \AA, boxcar smooth was 
applied to the data, to increase the apparent signal-to-noise ratio.  
The SiC 2B and LiF 1A {\it FUSE} channels containing the O VI doublet were used to determine the radial velocity solution
and exposure-to-exposure wavelength shifts were found to be negligible by measuring shifts in the geocoronal lines.
Radial velocity and line flux measurements for the brighter O VI $\lambda$ 1032 line were phased 
on the white dwarf spin period of 545.455 sec (Allan et al. 1996), but neither parameter produced an 
obvious modulation related to the spinning WD.  
Measurements of the O VI narrow emission-line components were phased on the orbital period.  The resulting
radial velocity curve is shown in Figure 3.  
We used the orbital ephemeris from Szkody et al. (2000) ($\phi$(HJD)=2,451,088.76164+0.1726196E)
in which they defined phase zero by a measured H$\alpha$ red-blue crossing (ie., the red-to-blue crossing
of the accretion disk/white dwarf), noting a zero phase difference of 0.23 between their 
observations and those of Haberl et al. (1994).  The zero phase error in extrapolating the ephemeris from Szkody et al. (2000) 
to our epoch is 0.07.
From our radial velocity results and by comparing our {\it FUSE}
radial velocity curve of the O VI narrow emission-lines to the optical emission-line radial velocities 
reported by Haberl et al. (1994; where their zero phase corresponds to blue-to-red crossing), we find 
that the O VI narrow emission-lines phase, and thus can be attributed to, formation at or near the white dwarf.
A $\chi^{2}$ fit of the radial velocity data was performed using a sinusoid of the form 
  \begin{equation}  V_{WD}(t) =  \gamma  + K_{WD} \sin{\left[  2\pi\frac{(t-T_{0})}{P} - 2\pi\phi\right]},   \end{equation}
with the orbital parameters as listed in Table 1, where
 $K_{WD}$ is the K velocity, P is the orbital period, T$_{0}$ and t are, respectively, the epoch and time, $\phi$ is the zero phase
difference, and $\gamma$ is
the system gamma velocity.  
It should be noted that although the source of the zero phase difference between Haberl et al. (1994) and 
Szkody et al. (2000) is unknown and a similar discrepancy between our {\it FUSE} data and Szkody et al. (2000) is possible, our
zero phase is within 0.01 of the value from Szkody et al. (2000). 
The mean value of our radial velocity measurements, 41.1$\pm$0.1 km s$^{-1}$, is in fair 
agreement with the systemic velocity of 31$\pm$2 km s$^{-1}$ reported by Haberl et al. (1994).  
The possible discrepancy in this {\it FUSE} velocity can be attributed to the use of a large aperture, resulting in 
velocities that can not be precisely placed on an absolute scale.
The difference might also be due to a bias imparted through comparison of optical disk emission lines
(Haberl et al. 1994) and FUV lines.
No obvious orbital correlation was seen in phasing radial velocity measurements of the FUV broad emission line 
components, although this was not a robust test due to the low signal-to-noise of these lines.

\section{Results}
The fit to the radial velocity measurements of the narrow O VI emission-line gives K$_{wd} = $2.5 $+/-$ 0.5 km s$^{-1}$.  
Assuming different mass values for the primary and secondary star, a probable range of inclination, $<$ 5$^{\circ}$, for 
V405 Aur can be estimated using K$_{wd}$ and Eq. (8) in Huber et al. (1998). 
The most likely theoretical value for the mass of the secondary star is 0.36 M$_{\odot}$ (Howell et al. 2001) however, we can not set any
meaningful constraints on the white dwarf mass. 
Further evidence for a low inclination, $<$ 5$^{\circ}$, comes from the small K$_{wd}$ amplitude and 
the single peaked nature of the broad optical and UV emission-lines which can be attributed to the accretion disk.
The broad {\it FUSE} emission-lines observed (C III $\lambda$1175 and O VI $\lambda$1032) are single peaked, as 
well as those seen in the optical (Szkody et al. 2000) and in an unpublished Hubble Space Telescope UV spectrum (Figure 4 \& 5).  

In AM Her, the narrow O VI emission-line components are attributed to 
irradiation of the secondary star, as the line velocities phase with the secondary star and have 
FWHM values far less than those of a typical accretion disk (Hutchings et al. 
2002).  The narrow O VI emission-lines (FWHM $\sim$70 km s$^{-1}$) we see in V405 Aur are narrower than typical accretion disks as well, 
but their orbital phasing identifies their source at or near the white dwarf, not the secondary star.  Since 
V405 Aur and AM Her have similar orbital 
periods and binary separations and possibly similar temperature WDs, the lack of irradiation 
producing narrow-lines in V405 Aur is likely due to shielding of the secondary 
star by the truncated accretion disk, or alternately, the ionization parameter (X-ray luminosity) could 
be too low as is the case with EX Hya (Mauche 1999) due to the larger accretion regions in IPs.  

Haberl et al. (1994) reported non-zero X-ray flux throughout the orbit of V405 Aur and detected modulations 
in the X-ray light curve on the white dwarf spin period.  We, however, do not see 
such a modulation in the O VI emission-lines, line flux, or radial velocities on the spin period.
A possible simple explanation could be that at such a low inclination, $\sim$5$^{\circ}$, no aspect
changes of the O VI emission region are seen by the observer.  In this scenario, the X-ray modulations would result from a 
partial self-eclipse of the magnetic pole(s), while the O VI emission region would form higher up the 
accretion column and not be eclipsed.  

The flux ratio of the narrow O 
VI emission-lines ($\lambda1032/\lambda1038$) remained nearly constant at 2.2$\pm$0.4 (Figure 6), indicating their 
formation in an 
optically thin gas (Mauche 1999).  This interpretation is complicated, however, by the 
lack of an observable continuum flux in the V405 Aur {\it FUSE} spectra.  It is possible that the {\it FUSE} spectra 
only detect the ``tips'' of the emission lines (especially O VI $\lambda$1038), making their ratio appear higher than it is.  Indeed, a 
standard T$_{eff}$ = 25,000 K, 0.55 M$_{\odot}$ DA white dwarf at the estimated 300 
pc distance (Haberl 1994) would produce a continuum flux of 3$\times$10$^{-14}$ 
ergs cm$^{-2}$ s$^{-1}$ \AA$^{-1}$ at 1000 \AA, well below the co-added {\it FUSE} flux 
detection limit (see Figure 1).  

By fitting a black body curve to the Haberl et al. (1994) optical spectrum plus unpublished {\it IUE} spectrum 
(LWP31948 and LWP31949 which show an apparent continuum roll-off near 2500 \AA) and making the assumption that
the continuum in dominated by the accretion disk flux in the optical and UV regions,
we estimate the accretion disk to have a maximum temperature of ~11,000 K.  At this temperature, the 
accretion disk is too cool to contribute to the 
continuum flux in the far-ultraviolet wavelength range.  The FWHM accretion disk 
velocity measured from the broad C III $\lambda$1175 \AA\ emission-line is $\sim$ 1100 km s$^{-1}$ in agreement 
with FWHM velocities obtained from optical data.
Accretion disk line widths for intermediate polars with truncated accretion disks range from 1400-1900 km s$^{-1}$ (Mauche 1999), 
indicating that V405 Aur contains a fairly typical truncated accretion disk in which the inner disk is destroyed by 
material threading along the magnetic field onto the white dwarf (see Belle et al. 2003).

\section{Conclusions}

{\it FUSE} spectra of the magnetic cataclysmic variable V405 Aur show the system to 
be an intermediate polar having a low inclination.    Radial velocity measurements of the 
narrow O VI emission-lines reveal their source to be on or near the magnetic poles of the white dwarf, not irradiation of 
the secondary as is the case with AM Her.  Since the narrow O VI emission-lines show 
neither radial velocity nor line flux variations on the white dwarf spin period, 
their origin appears to be higher up in the near WD accretion columns, unlike the X-rays
which may be spin modulated due to self-eclipse of the pole(s).
Given the difficulty of observing the white dwarf in IP systems, far-ultraviolet 
emission-lines might prove to be the only method of more-or-less direct measurement of the compact object.  

V405 Aur, PQ Gem, RXJ0512, and RXJ0757 all have magnetic field 
strengths at the low end of the distribution in polars and all four have fairly long orbital periods.
The strength of the magnetic field in V405 Aur has been estimated to be $>$ 5 MG by the flux 
ratio of hard bremsstrahlung to soft black body in the {\it ROSAT} band (Haberl \& Motch 
1995).  This value is consistent with the strongest field strengths of IPs as well as the lowest field strength  
polars such as AM Her which has a field strength of 14.5 MG (Bailey et al. 1991).  
These systems are likely to become AM Her-like polars as their orbital and spin periods synchronize in an astrophysically 
short time.  With the average magnetic field strength of polars being much stronger than those in 
typical IPs, including these four possible transition objects, the question then remains; where are 
the magnetically strong IPs which will 
become magnetically strong polars?  Perhaps they undergo a rapid pre-polar to polar evolution becoming 
polars essentially at the same time they become mass transferring CVs.

\newpage

\acknowledgments
This research was partially supported by a NASA/FUSE research grant to SBH.
The {\it HST} data presented in this paper were obtained from the
Multimission Archive at the Space Telescope Science Institute (MAST).
STScI is operated by the Association of Universities for Research in
Astronomy, Inc., under NASA contract NAS5-26555. Support for MAST for
non-HST data is provided by the NASA Office of Space Science via grant
NAG5-7584 and by other grants and contracts.  We also thank the anonymous 
referee for useful comments.

\begin{planotable} {lll}
\tablenum{1}
\tablewidth{43pc}
\tablecaption{V405 Aur System Parameters}
\vspace{3mm}
\tablehead{
\colhead{Parameter}         &
\colhead{Value}          &
\colhead{Reference}    }
\startdata
Orbital Period           & 4.15 hours                       &  Haberl et al. (1994)    \nl
WD Spin Period           & 545.4565(8) sec                  &  Allan et al. (1996)     \nl
Estimated Distance       & 300-310 pc                       &  Haberl et al. (1994)    \nl
Estimated Magnetic Field Strength &  $>$ 5 MG               &  Haberl \& Motch  (1995) \nl
Radial velocity ephemeris T$_{0}$& HJD 2,451,088.76164+0.1726196E & Szkody et al. (2000)\nl
Zero phase difference $\phi$     &  -0.01$\pm$0.07          &  this work               \nl
Inclination              &  $<$ 5 degrees                   &  this work               \nl
$\gamma$ velocity        &  +41.1$\pm$0.1 km s$^{-1}$       &  this work               \nl
$K_{WD}$ Velocity        & 2.5$\pm$ 0.5 km s$^{-1}$         &  this work               
\enddata
\end{planotable}

\begin{planotable} {ccccc}
\tablenum{1}
\tablewidth{43pc}
\tablecaption{Log of {\it FUSE} Spectroscopic Observations}
\vspace{3mm}
\tablehead{
\colhead{Exposure Name}         &
\colhead{MJD Start}         &
\colhead{MJD End}         &
\colhead{Exposure Time}         &
\colhead{Binary Orbital Phase}          \\
\colhead{}         &
\colhead{(JD-2,450,000)}         &
\colhead{(JD-2,450,000)}         &
\colhead{(seconds)}         &
\colhead{}          }
\startdata
D0800101001 &	2913.08758 &	2913.11561 &	2422 &	0.56 \\
D0800101901 &	2913.11590 &	2913.12695 &	 955 &	0.67 \\
D0800101003 &	2913.15425 &	2913.18488 &	2647 &	0.95 \\
D0800101902 &	2913.18521 &	2913.19673 &	 996 &	0.07 \\
D0800101004 &	2913.20171 &	2913.20446 &	 238 &	0.14 \\
D0800101005 &	2913.22638 &	2913.25415 &	2400 &	0.36 \\
D0800101903 &	2913.25447 &	2913.26647 &	1037 &	0.48 \\
D0800101006 &	2913.27255 &	2913.27730 &	 411 &	0.56 \\
D0800101007 &	2913.29834 &	2913.32342 &	2167 &	0.77 \\
D0800101904 &	2913.32373 &	2913.33613 &	1072 &	0.88 \\
D0800101008 &	2913.33909 &	2913.35157 &	1079 &	0.97 \\
D0800101009 &	2913.37021 &	2913.39269 &	1943 &	0.18 \\
D0800101905 &	2913.39300 &	2913.40578 &	1105 &	0.28 \\
D0800101010 &	2913.40789 &	2913.42558 &	1528 &	0.38 \\
D0800101011 &	2913.44204 &	2913.46196 &	1722 &	0.59 \\
D0800101906 &	2913.46225 &	2913.47541 &	1138 &	0.68 \\
D0800101012 &	2913.48132 &	2913.49967 &	1586 &	0.81 \\
D0800101013 &	2913.51362 &	2913.53124 &	1522 &	0.99 \\
D0800101907 &	2913.53155 &	2913.54502 &	1164 &	0.09 \\
D0800101014 &	2913.55101 &	2913.59935 &	4177 &	0.30 
\enddata
\tablecomments{The exposure names beginning with D08001019 were
taken during an occultation.}
\end{planotable}

\begin{figure}
        \includegraphics[angle=90,scale=.50]{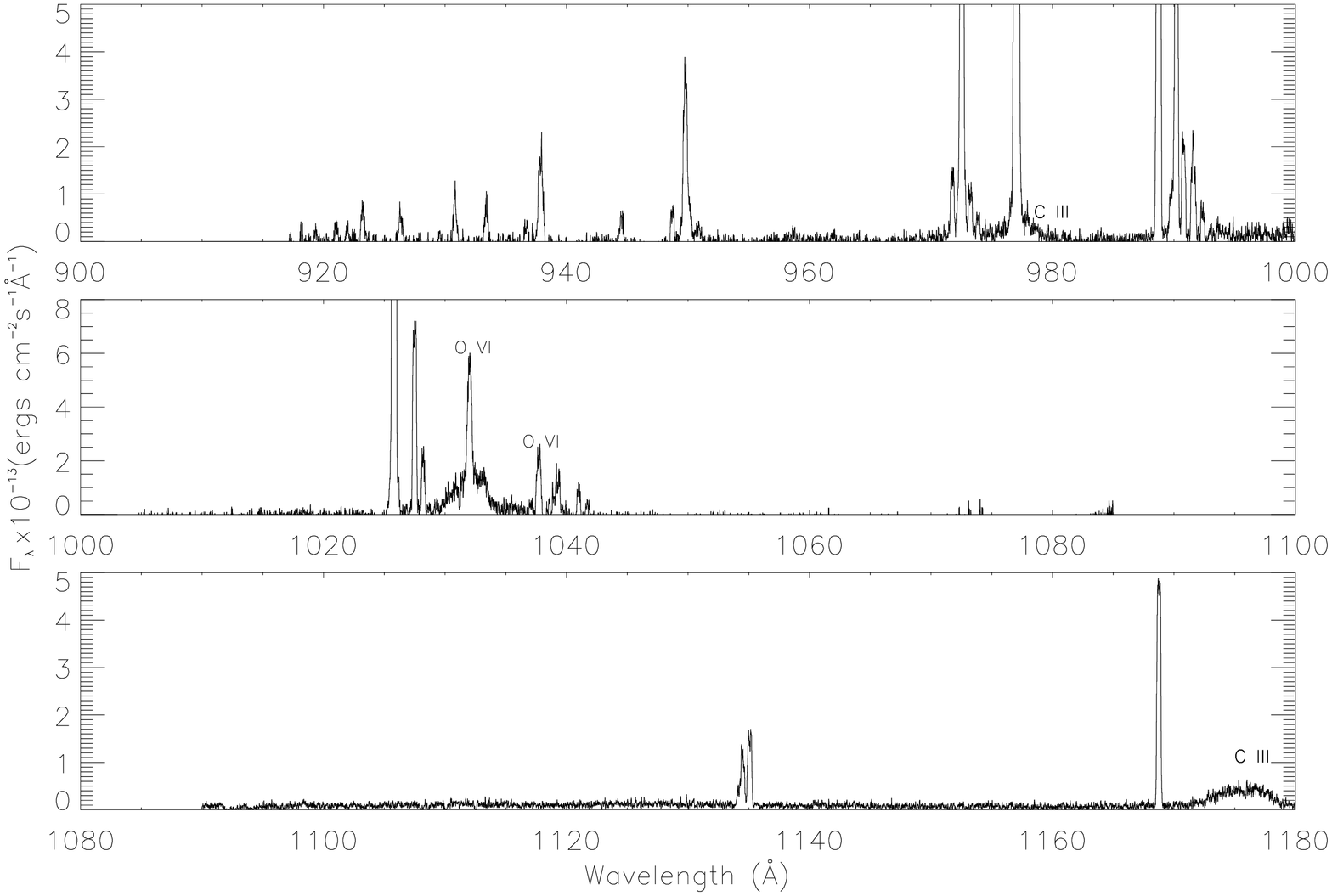}
        \caption{Coadded {\it FUSE} spectra, smoothed over 0.04\AA, showing broad and narrow emission-lines of O VI and broad C III emission lines. All other lines are geocoronal.}

       \includegraphics[angle=90,scale=.50]{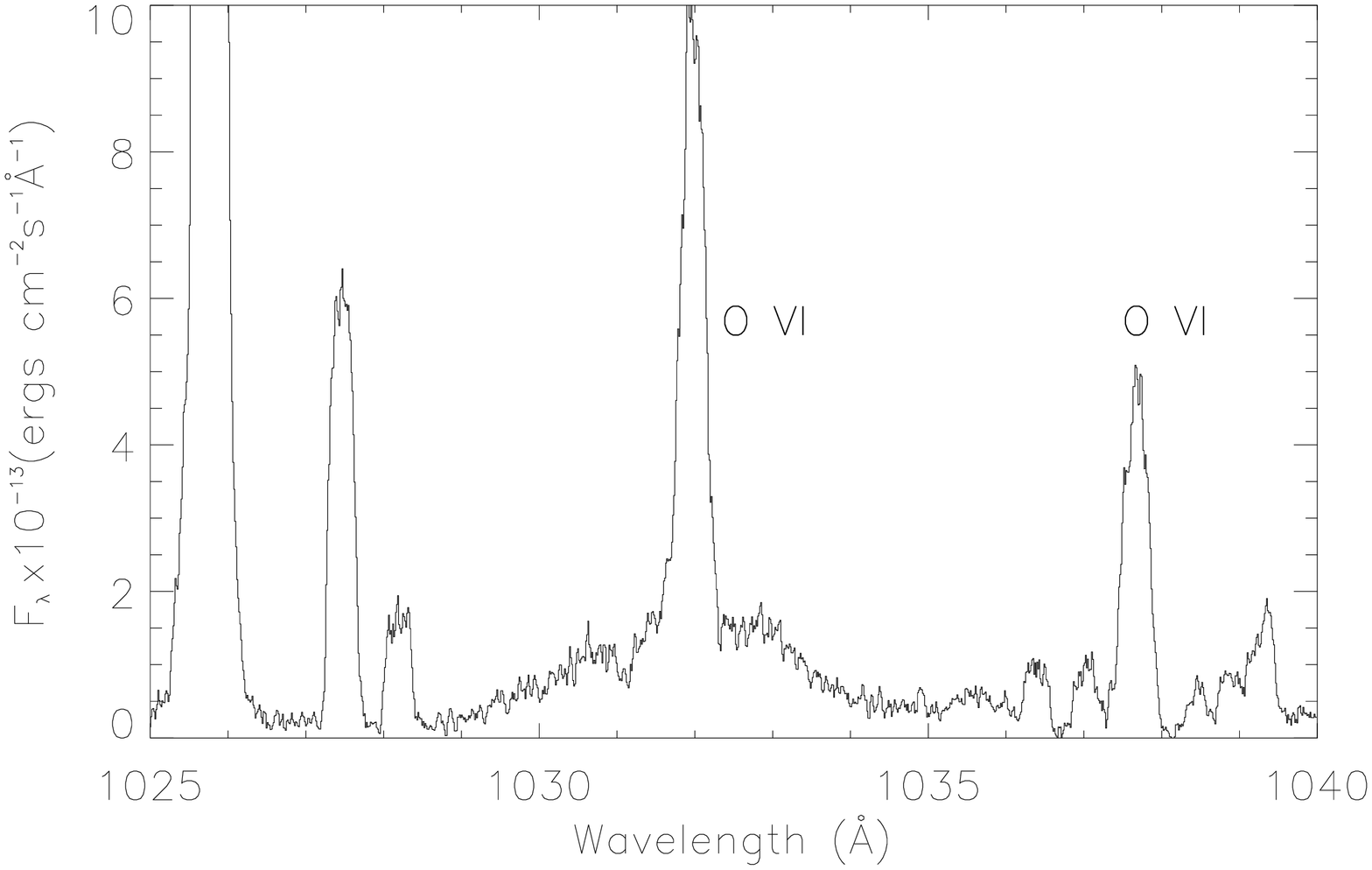}
       \caption{{\it FUSE} Spectra, smoothed over 0.04\AA, showing resolved O VI $\lambda\lambda$ 1032, 1038 \AA\ emission-lines.  Note the broad component of the O VI $\lambda$1032 and the lack of continuum flux. All other lines are geocoronal.}
\end{figure}

\begin{figure}
       \includegraphics[angle=90,scale=.50]{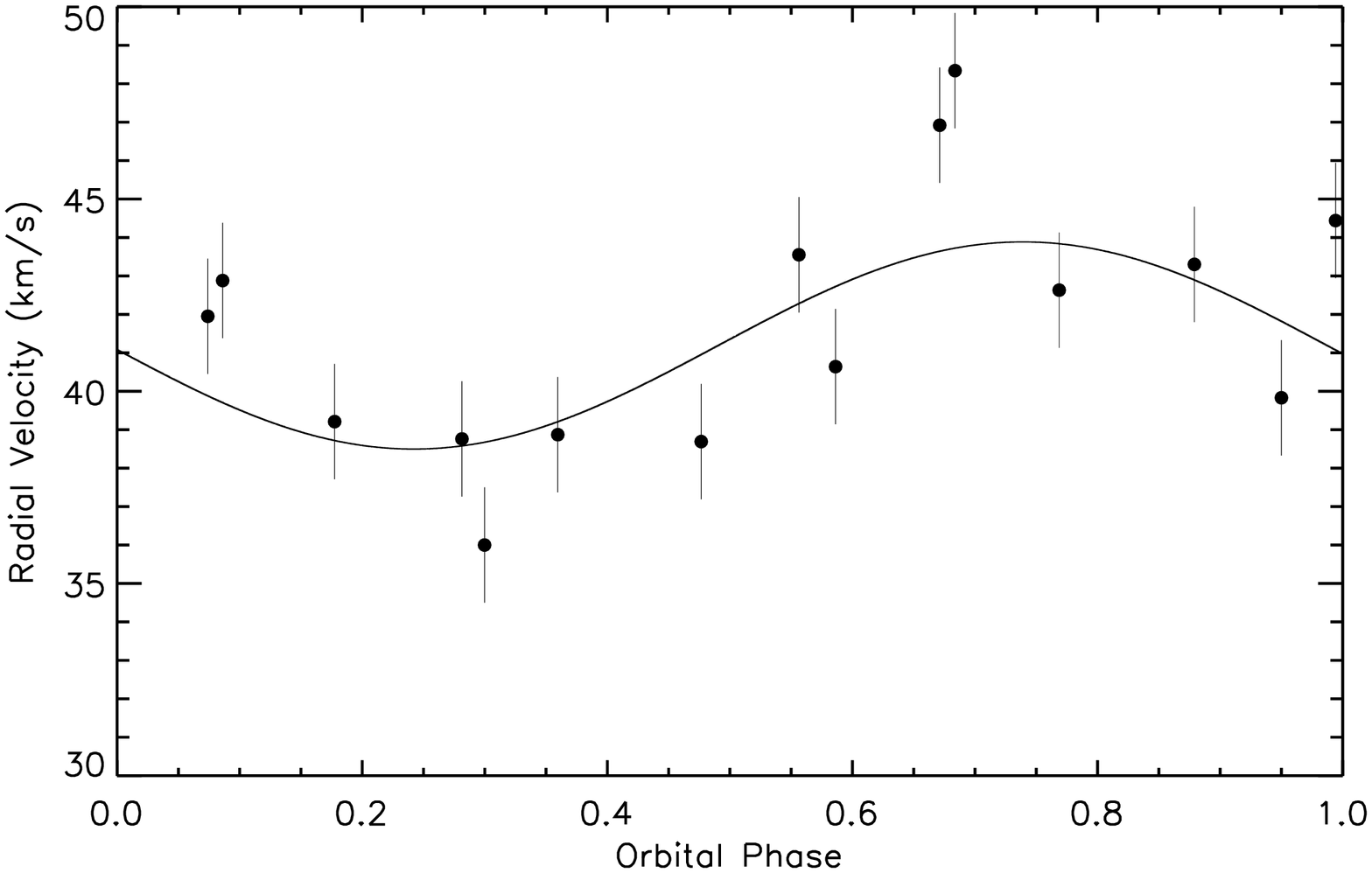}
       \caption{Radial Velocity curve of the O VI 1032 narrow emission-line phased on the orbital period.  The
red-to-blue crossing of the fit is at $\phi$=0.01 indicating correspondence with the motion of the white dwarf.}

       \includegraphics[angle=90,scale=.45]{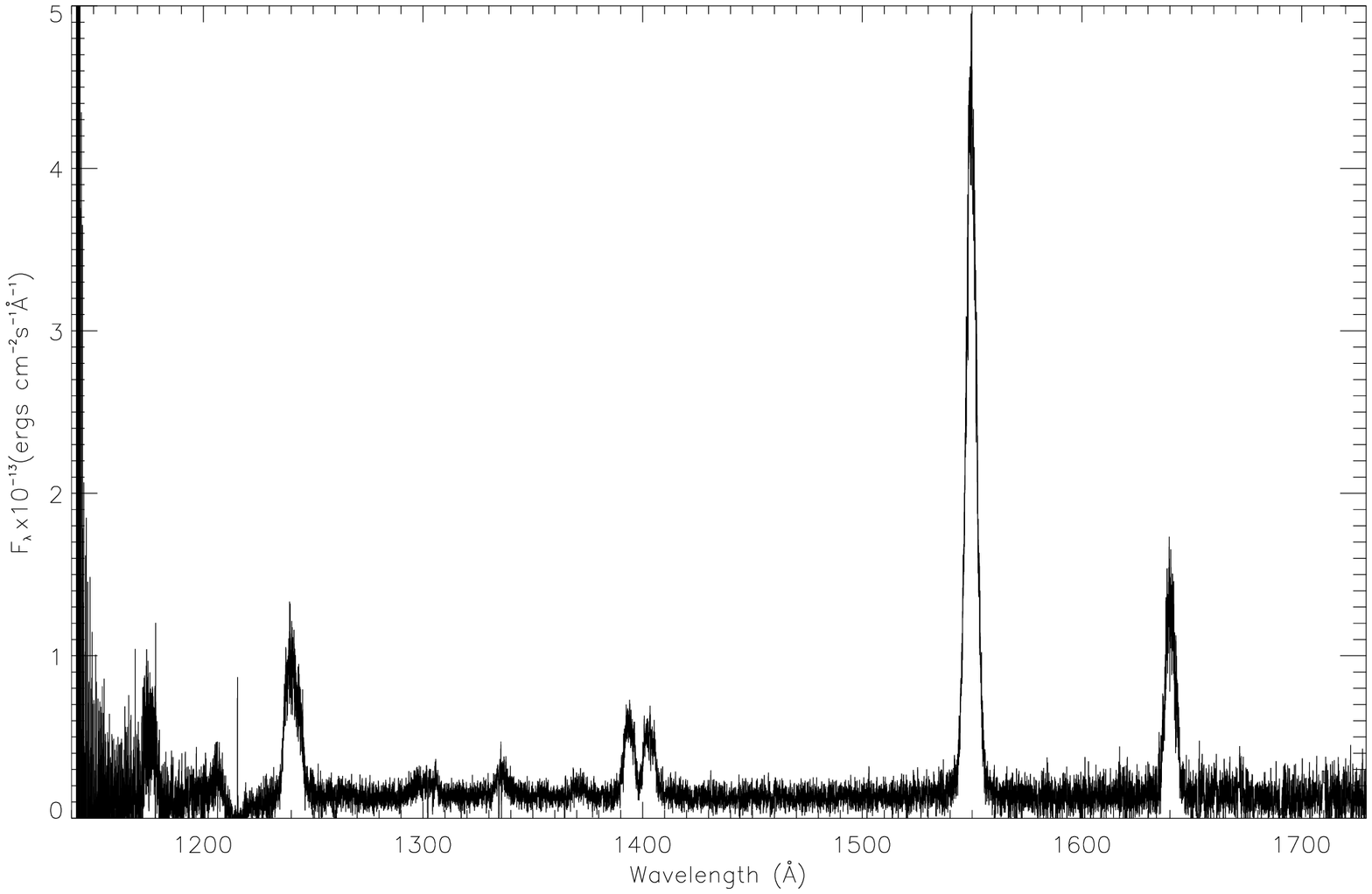}
       \caption{Hubble Space Telescope spectrum of V405 Aur showing N V, Si IV, C IV, and He II emission lines.  This unpublished spectrum was obtained from the Multimission Archive at the Space Telescope Science Institute.}

\end{figure}

\begin{figure}
       \includegraphics[angle=90,scale=.45]{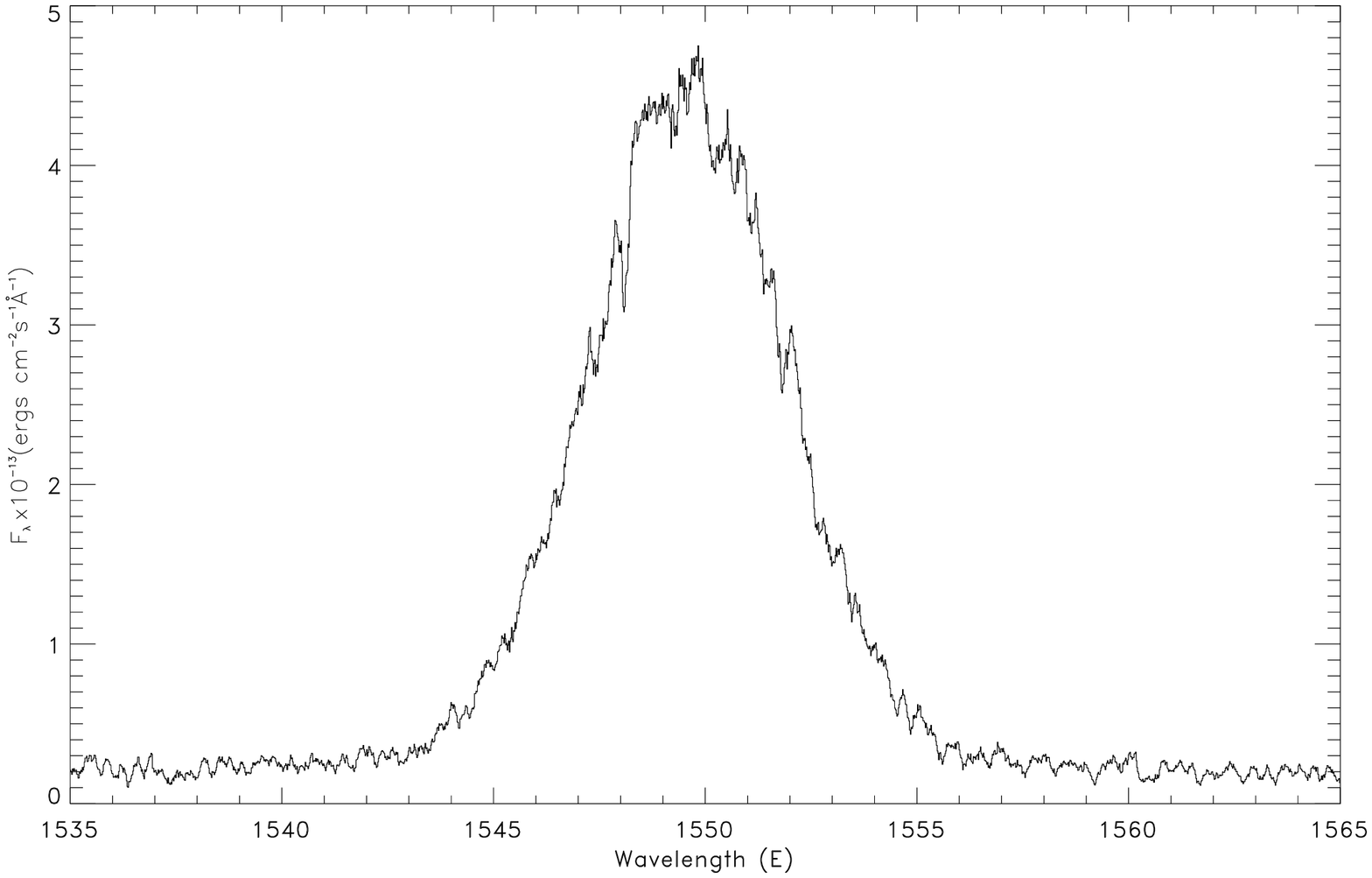}
       \caption{C IV doublet emission line profile from the Figure 4 spectrum having a FWHM of 5.4 \AA.  The single peaked nature of the UV emission lines is additional evidence for a low binary inclination.}

       \includegraphics[angle=90,scale=.50]{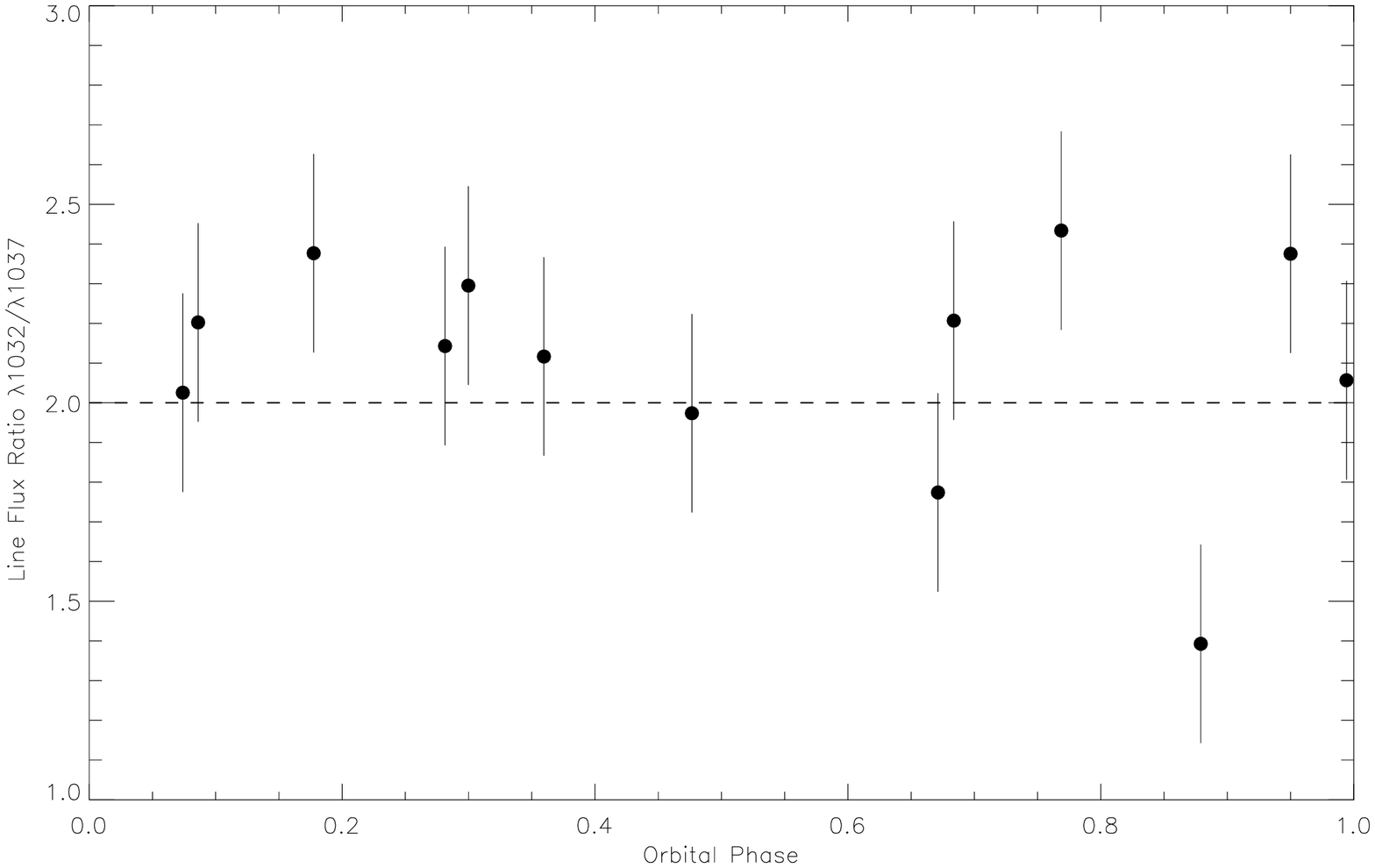}
       \caption{Line flux ratio of O VI $\lambda$1032/$\lambda$1038 vs. orbital phase.  A value of 2.0 indicates an origin in optically thin gas.}

\end{figure}

\end{document}